\documentstyle[aaspp4]{article}
\begin{document}


\title{Protostellar Collapse with Various Metallicities}

\author{Kazuyuki Omukai}
\affil{Department of Physics, Kyoto University, Kyoto 606-8502, Japan}


\begin{abstract}
The thermal and chemical evolution of gravitationally collapsing
protostellar clouds is investigated, focusing attention on their
dependence on metallicity.
Calculations are carried out for a range of metallicities spanning the
local interstellar value to zero.
During the time when clouds are transparent to continuous radiation, the
temperatures are higher for those with lower metallicity, reflecting
lower radiative ability.  
However, once the clouds become opaque, in the course of the adiabatic
contraction of the transient cores, their evolutionary trajectories in
the density-temperature plane converge to a unique curve that is
determined by only physical constants.  
The trajectories coincide with each other thereafter.
Consequently, the size of the stellar core at the formation is the same
regardless of the gas composition of the parent cloud.  
\end{abstract}

\keywords{ISM: clouds --- ISM: molecules --- molecular processes ---
  stars: formation --- stars: Population II } 

\newpage
\section{Introduction}
In the standard theory of cosmic structure formation, galaxies are
believed to have originated from primeval density fluctuations. 
Many authors have been studying the galaxy formation process in this
context and have made remarkable progress in understanding such
processes as the gravitational growth
of the density fluctuation, its decoupling from the cosmic expansion and
virialization, and the subsequent radiative cooling of the baryonic gas
(e.g., Padmanabhan 1993).
However, processes after the cooling of baryonic gas, namely, the
transformation of a pregalactic cloud into a cluster of stars, are 
relatively poorly known.
It is inevitably necessary to investigate them in studying the galaxy
formation process since the very existence of stellar components is one
distinct feature of galaxies.   
These processes can be viewed as successive fragmentations of a cloud into
fragments, or in other words, protostellar clouds, and their contraction
into stars (e.g., Hayashi 1984).

Whether a cloud fragments or collapses into a single object depends on the 
the central flatness, that is, the axial ratio of the isodensity contour
in the central region (Tsuribe \& Inutsuka 1999).
In particular, warm clouds do not fragment before adiabatic core
formation, whereas clumps sufficiently more massive than the Jeans mass
first collapse disklike and next fragment most likely into filamentary
clouds (Miyama, Narita, \& Hayashi 1987a, 1987b). 
The filamentary clouds fragment again into protostellar cloud cores
after further contraction.
Uehara et al. (1997) investigated the gravitational collapse of
metal-free filamentary clouds using one-zone approximation and found that the
minimum mass of fragments, or protostellar cores, is essentially
Chandrasekhar mass, i.e., $\sim 1M_{\sun}$. 
Nakamura \& Umemura (1999) confirmed Uehara et al. (1997)'s
result by performing one-dimensional hydrodynamical calculations. 
In this paper, we will discuss the collapse of protostellar cores into
stars.

The distinction between the star formation in the galaxy formation epoch and
that in present-day star-forming regions mainly resides in the
difference of the composition of gas that stars are made from.
Metallicity, and therefore the gas to dust ratio, is lower for 
earlier star formation.
In present-day star forming regions, protostellar clouds remain roughly 10K in
a wide range of densities owing to efficient dust emission (e.g.,
Hayashi \& Nakano 1965) in the course of contraction.

On the other hand, many authors have studied the thermal and chemical
evolution of collapsing primordial clouds with one-zone approximation 
 (e.g., Matsuda, Sato, \& Takeda 1969; Yoneyama 1972; Calberg 1981;
 Palla, Salpeter, \& Stahler 1983; Izotov \& Kolesnik 1984; Lahav 1986;
 Puy \& Signore 1997) or by hydrodynamical calculations (Matsuda et
 al. 1969; Villere \& Bodenheimer 1987; Haiman, Thoul, \& Loeb
 1996; Omukai \& Nishi 1998; Oliveira et al. 1998).  
The collapse of metal-free protostellar clouds, which is
relevant for the first star formation in the universe, is induced
by radiative cooling owing to molecular hydrogen lines, and their
temperatures are about 1000K. 

Above are two extremes in composition of protostellar gas, namely,
the former for the high-metallicity end and the latter for the
low-metallicity end. 

Then, how does the collapse of slightly metal-polluted clouds proceed?
Even in the galaxy formation epoch, significant metal enrichment can
occur since the lifetime of massive stars ($\sim
10^{6}$ yr) is shorter than the free-fall time of pregalactic clouds
($\sim 10^{8}$ yr). 
In fact, both Ly$\alpha$ forest clouds and Population II stars are
slightly polluted by heavy elements.  
Therefore, it is important to study star formation from slightly
metal-polluted gas. 
In this paper, we aim to fill the gap between the protostellar collapse
of primordial and present-day interstellar gas.  

This topic has been investigated in the pioneering works of Low \&
Lynden-Bell (1976) and Yoshii \& Sabano (1980).  
However, the following points remain to be fulfilled;
in their works,
(i) chemical reactions had not been solved, 
(ii) molecular coolants, which become important at high densities,
had not been considered, 
and (iii) the evolution of clouds as dense as the stellar density
($>10^{22} {\rm cm^{-3}}$) had not been studied. 
With these points in mind, we investigate in this paper the thermal and
chemical evolution of collapsing spherical clouds as a function of
metallicity.

The outline of this paper is as follows.
In \S 2, we describe the method of our calculations. 
In \S 3, results of our calculations are presented.
We summarize our work in \S 4.

\section{Model}
\subsection{Basic Equations}
We consider a spherical cloud with mean metallicity $Z$.
The helium concentration is assumed to be $y_{\rm He}=0.0972$ for all
clouds.\footnote{The concentration of element X is defined by
\begin{equation}
y_{X}=n_{X}/n,
\end{equation}
where $n$ and $n_{X}$ are the number densities of hydrogen nuclei and 
nuclei of element X.
Similarly, we write for each atomic, molecular, or ionic species 
\begin{equation}
y(x)=n(x)/n,
\end{equation}
where $n(x)$ is the number density of species $x$.
Note $y({\rm H_2})=1/2$ for fully molecular gas.    
}
We assume that the dust-to-gas ratio is proportional to the mean
metallicity, or in other words, that a fixed fraction of the heavy elements
in the interstellar medium condenses into dust grains.
We adopt the Pollack et al. (1994) model of grains in molecular clouds
(see \S 2.2.2).
For the case of local interstellar clouds ($Z \simeq 1 Z_{\sun}$),
the mass fraction of grain we used is $0.934 \times 10^{-2}$ in the
lowest temperature regime, and the gas-phase elemental abundances are
$y_{\rm C}=0.927 \times 10^{-4}$, and $y_{\rm O}=3.568 \times 10^{-4}$,
respectively, which correspond to 46\% of oxygen and 72\% of carbon
depleted onto grains.
These values are reduced proportionally for lower metallicity cases.  
We normalize the mean metallicity (and therefore also the dust-to-gas
ratio and gas-phase metallicity in our model) relative to the local
interstellar values $Z_{\rm local}$ and denote 
relative metallicity by $z \equiv Z/Z_{\rm local}$. 
 
We calculate the time evolution of the central density, temperature, and
chemical composition of the collapsing core.

First, we assume that the dynamics is described by the free-fall
relation,
\begin{equation}
\frac{d \rho}{dt}=\frac{\rho}{t_{\rm ff}},
\end{equation}
where $\rho$ is the density in the central region and the free-fall time 
is
\begin{equation}
t_{\rm ff}\equiv \sqrt{\frac{3 \pi}{32 G \rho}}.
\end{equation}

The thermal evolution is followed by solving the energy equation
\begin{equation}
\frac{d e}{dt}=-p \frac{d}{dt} (\frac{1}{\rho})-{\cal L}^{(\rm net)},
\end{equation}
where
\begin{equation}
e=\frac{1}{\gamma_{\rm ad}-1} \frac{k  T}{\mu m_{\rm H}}
\end{equation}
is the specific internal energy, 
\begin{equation}
p=\frac{\rho k T}{\mu m_{\rm H}}
\end{equation}
is the pressure for an ideal gas, $\gamma_{\rm ad}$ is the adiabatic
exponent, $T$ is the temperature, $\mu$ is the mean molecular weight,
$m_{\rm H}$ is the mass of hydrogen nucleus, 
and ${\cal L}^{(\rm net)}$ is the net energy loss rate per unit mass
(see \S 2.2).

We neglect any effect owing to rotation or magnetic fields for
simplicity.
In this case, the actual collapse is expected to proceed like the
Penston-Larson similarity solution (Penston 1969; Larson 1969). 
According to this solution, the cloud consists of two parts, that is, the
central core region, which has flat density distribution, and the
envelope, where the density decreases outward as $\propto r^{-2}$.
The size of the central flat region is roughly given by the local
Jeans length $\lambda _{\rm J}=\pi c_{\rm s}/\sqrt{G \rho}$ in the
core. 
Since we focus on the evolution of the central region, optical depth is
estimated by that across one local Jeans length:
\begin{equation}
\tau_{\nu}=\kappa_{\nu}\rho \lambda_{\rm J}.
\end{equation}
This formulation of Jeans length shielding is the same procedure as
that of Low \& Lynden-Bell (1976) and Silk (1977), although they
introduced it as a result of successive fragmentation into Jeans mass
clouds, not because of the Penston-Larson collapse. 
      
\subsection{Cooling/Heating processes}
In addition to compressional heating, we include cooling owing
to (i) atomic and molecular line radiation,
(ii) energy transfer between the gas and the dust grains, which will be
emitted as infrared radiation from the grains, 
(iii) continuous radiation from the gas, 
and (iv) cooling and heating associated with chemical reactions.
Then
\begin{equation}
{\cal L}^{(\rm net)}={\cal L}_{\rm line}+{\cal L}_{\rm gr}+{\cal L}_{\rm
  cont}+{\cal L}_{\rm chem}, 
\end{equation}
where the terms in the right-hand side of the equation correspond to the
cooling rates owing to the processes (i)-(iv) above in the same order.  
Although we treat clouds with metallicity up to $\simeq 1 Z_{\sun}$, we
are concerned mainly with the lower metallicity environments that are
relevant to the epoch of galaxy formation.
In this paper, we do not try to reproduce the present Galactic
environment and neglect any external radiation (e.g., UV radiation,
cosmic rays, cosmic background radiation, etc.) for simplicity.  
Then, the considered heating sources are the compressional work and
H$_2$ formation. 
As will be mentioned below, lower metallicity clouds tend to be warmer.
Therefore, the compressional heating, which is proportional to the
temperature of the cloud, becomes more important if the external heating 
rate is constant.
\subsubsection{Line Cooling}
As line cooling agents, we include atomic fine-structure transitions of
CI, CII, and OI and molecular rovibrational transitions of H$_2$, CO,
OH, and H$_2$O: 
\begin{equation}
{\cal L}_{\rm line}={\cal L}_{\rm CI}+{\cal L}_{\rm CII}+{\cal
  L}_{\rm OI}+{\cal L}_{\rm H_2}+{\cal L}_{\rm CO}+{\cal
  L}_{\rm OH}+{\cal L}_{\rm H_2O}.
\end{equation}
The line cooling rate of species $x$ is given by
\begin{equation}
{\cal L}_{x}=\frac{1}{\rho}\sum_{(i \rightarrow j)} n(x,i) A_{ij}
\epsilon_{ij} h \nu_{ij},
\end{equation}
where $n(x,i)$ is the population density of species $x$ in level
$i$, $A_{ij}$ is the spontaneous transition probability, $\epsilon_{ij}$
is the escape probability, and $h \nu_{ij}$ is the energy difference
between levels $i$ and $j$. 
The population density $n(x,i)$ is obtained from a solution of the equation of
statistical equilibrium
\begin{equation}
n(x,i)\sum_{j \neq i}^{n} R_{ij}=\sum_{j \neq i}^{n} n(x,j) R_{ji},
\end{equation}
where $n$ is the total number of lines included and 
\begin{equation}
R_{ij} = \left\{
\begin{array}{ll}
 A_{ij} \epsilon_{ij}+C_{ij}
& \mbox{for $i>j$} \\
 C_{ij}
& \mbox{for $i<j$},
\end{array}
\right.
\end{equation}
ignoring the external radiation.
Here $C_{ij}$ is the collisional rate from level $i$ to level $j$.
Following Takahashi, Hollenbach, \& Silk (1983), we use the escape
probability for the case that the velocity is proportional to the radius
in the central region:
\begin{equation}
\epsilon_{ij}= \left( \frac{1-{\rm e}^{-\tau_{ij}}}{\tau_{ij}} \right)
{\rm e}^{-\tau_{\rm cont}}. 
\end{equation}

The optical depth averaged over the line and the continuum optical depth
are given by   
\begin{equation}
\tau_{ij}=\frac{A_{ij} c^{3}}{8 \pi
  \nu_{ij}^{3} \eta_{\rm T}}[n(x,j)g_{i}/g_{j}-n(x,i)]l_{\rm sh}/ (2
  \Delta v_{\rm D}),
\end{equation}
and 
\begin{equation}
\tau_{\rm cont}=(\kappa_{\rm gr}+\kappa_{\rm gas})\rho \lambda_{\rm J},
\end{equation}
respectively,
where $\eta_{\rm T}$ is the multiplicity factor, $g_{i}$ is the
statistical weight of level $i$, $\Delta v_{\rm D}$ is the velocity
dispersion, and $l_{\rm sh}$ is the shielding length (Takahashi et
al. 1983):
\begin{equation}
l_{\rm sh}={\rm min}(\Delta s_{\rm th}, \lambda_{\rm J}); ~~~~
\Delta s_{\rm th}=2 \Delta v_{\rm D}/(\frac{dv}{dr})= 6 \Delta v_{\rm D}
t_{\rm ff},
\end{equation}
where we have used the relation for the homogeneous collapse $v=r/3
t_{\rm ff}$ and, in the case of the
small velocity gradient, have assumed Jeans length shielding as
introduced in \S 2. 
As a source of the velocity dispersion, we consider only the thermal
motion of atoms and neglect microturbulent motions.
Then 
\begin{equation}
\Delta v_{\rm D}=\sqrt{\frac{2 k T}{\mu_{x} m_{\rm H}}},
\end{equation}
where $\mu_{x}$ is the molecular weight of species $x$.
The continuum Planck mean opacity of dust $\kappa_{\rm gr}$ and that
of gas $\kappa_{\rm gas}$ are adopted from Pollack et al.(1994) and Lenzuni,
Chernoff, \& Salpeter (1991), respectively (see \S 2.2.2 and 2.2.3).

In general, the level population and the escape probabilities depend on
each other. 
We then need an iterative procedure to find their consistent solution
except for the case of H$_2$.
For H$_2$ lines, we need not iterate to find the consistent set of the
level populations and the escape probabilities, since in our
calculations, clouds become opaque to H$_2$ lines only at the density
exceeding the critical density, and in such a case, molecules populate
following a Boltzmann distribution regardless of the escape probabilities.

The related parameters of transitions are given in
Hollenbach \& McKee (1989) for CI, CII, and OI, 
in Hollenbach \& McKee (1979;1989) for OH and H$_2$O
\footnote{The OH and H$_2$O collision cross sections and H$_2$O
  multiplicity factor are given in Hollenbach \& McKee (1989).
  We adopted other parameters from Hollenbach \& McKee (1979).},  
and in McKee et al. (1982) for CO.
We computed the population of H$_2$ following the procedure of Hollenbach
\& McKee (1979) using the collision coefficient given in Hollenbach \& McKee
(1989).
For H$_2$, we considered the first three vibrational states ($v=0-2$) with
rotational levels up to $J=20$ in each vibrational state.
We take into account only the ground vibrational level for other
molecular lines since cooling rates owing to vibrational transitions are
nearly always dominated by grain cooling or H$_2$ cooling (Hollenbach
\& McKee 1979). 

\subsubsection{Gas-Grain Heat Transfer}
We adopt the Pollack et al. (1994) model of grains in molecular clouds.
In their model, silicates, organics, troilite, metallic iron, and water
ice constitute the most abundant grain species and the dominant sources 
of opacity in molecular cloud cores.
The Planck mean opacity due to the grains in the case of mean
metallicity $Z=Z_{\rm local}$ is 
\begin{equation}
\kappa_{\rm gr}=4.0 \times 10^{-4} T_{\rm gr}^{2} ~{\rm cm^{2}~g^{-1}},
\end{equation}
in $T_{\rm gr} \lesssim 50 {\rm K}$, where $T_{\rm gr}$ is the effective 
grain temperature (see below). 
The mass fraction of grain is $0.934 \times 10^{-2}$ below the water
vaporization temperature (100-200 K, depending on the density), and that
of each component; water ice, volatile organics, refractory organics,
troilite, orthopyroxene, olivine and metallic iron are $1.19 \times
10^{-3}, 6.02 \times 10^{-4}, 3.53 \times 10^{-3}, 5.69 \times 10^{-4},
7.33 \times 10^{-4}, 2.51 \times 10^{-3}$, and $2.53 \times 10^{-4}$,
respectively (from Table 2 of Pollack et al. 1994).
For the grain size distribution
\begin{equation}
n(a)\propto \left\{
\begin{array}{ll}
a^{-3.5} & \mbox{($0.005{\rm \mu m}<a<1 {\rm \mu m}$)} \\
a^{-5.5} & \mbox{($1{\rm \mu m}<a<5 {\rm \mu m}$)}
\end{array}
\right.
\end{equation}
(Pollack et al. 1994) that is based on that derived by Mathis, Rumpl,
\& Nordsieck (1977), 
the energy transfer rate from the gas to the dust grains per unit mass 
is given by  
\begin{equation}
{\cal L}_{\rm gr}=1.1 \times 10^{5} n (\frac{f_{\rm
    gr}}{\rho_{\rm gr}}) 
(\frac{T}{1000 {\rm K}})^{1/2} [1-0.8 {\rm exp}(-75 {\rm K}/T)](T-T_{\rm gr})
\end{equation}
(Hollenbach \& McKee 1979),
where $T_{\rm gr}$ is an effective grain temperature, which is determined 
by the energy balance for dust grains;
\begin{equation}
{\cal L}_{\rm gr}=4 \sigma T_{\rm gr}^{4} \kappa_{\rm gr} \beta_{\rm cont},  
\end{equation}
where $\kappa_{\rm gr}$ is the Planck mean opacity owing to the dust
grains and $f_{\rm gr}/\rho_{\rm gr}$ is the total volume of dust grains
per unit mass of gas and $f_{\rm gr}/\rho_{\rm gr}=5.3 \times 10^{-3}$ at the
lowest temperatures (i.e., below the water ice vaporization
temperature).
These quantities are taken from tables in Pollack et al.(1994).

The continuum energy transport rate $\beta_{\rm cont}$ decreases as
$\tau_{\rm cont}^{-2}$ owing to radiative diffusion in the optically
thick case (e.g., Masunaga et al. 1998).
Then, 
\begin{equation}
\beta_{\rm cont}={\rm min}(1,\tau_{\rm cont}^{-2}).
\end{equation}

\subsubsection{Continuum of Gas}
Using the Planck mean opacity of gas $\kappa_{\rm gas}$, the cooling
rate owing to the gas continuum is given by
\begin{equation}
{\cal L}_{\rm cont}=4 \sigma T^{4} \kappa_{\rm gas}
\beta_{\rm cont}.   
\end{equation}
We take the continuum Planck mean opacity for
metal-free gas from Lenzuni et al. (1991), which includes
all the important continuum processes, specifically, bound-free
absorption by 
${\rm H^{0}}$ and ${\rm H^{-}}$; free-free absorption by ${\rm H^{0}}$,
${\rm H^{-}}$, ${\rm H_{2}}$, ${\rm H_{2}^{-}}$, ${\rm H_{2}^{+}}$,
${\rm H_{3}^{+}}$, ${\rm He^{0}}$, ${\rm He^{-}}$; photodissociation of
${\rm H_{2}}$, and ${\rm H_{2}^{+}}$ by thermal radiation; Rayleigh
scattering by ${\rm H^{0}}$, ${\rm H_{2}}$, ${\rm He^{0}}$; 
Thomson scattering by $e^{-}$; and
collision-induced absorption by ${\rm H_{2}}$ due to collisions with
${\rm H_{2}}$, ${\rm He^{0}}$, and ${\rm H^{0}}$.
Among the above processes, the most important cooling mechanism is H$_2$
collision-induced continuum, which dominates the cooling at about $\sim
10^{16} {\rm cm^{-3}}$ for low metallicity (i.e., $<10^{-6} Z_{\sun}$)
clouds.

\subsubsection{Chemical Cooling/Heating}
Following Hollenbach \& McKee (1979), we assume the heat deposited per a
formed molecular hydrogen as 
$0.2+4.2(1+n_{\rm cr}/n)^{-1}~{\rm eV}$
for H$_2$ formation on grain surfaces (reaction 23 in Appendix),
$3.53(1+n_{\rm cr}/n)^{-1}~{\rm eV}$
for H$_2$ formation by H$^{-}$ process (reaction 8), 
$1.83(1+n_{\rm cr}/n)^{-1}~{\rm eV}$
for H$_2$ formation by H$_2^{+}$ process (reaction 10), and 
$4.48(1+n_{\rm cr}/n)^{-1}~{\rm eV}$
for H$_2$ formation by the three-body reactions (reactions 19 and 20), 
where 
\begin{equation}
  n_{\rm cr}=\frac{10^{6} T^{-1/2}}
{1.6y({\rm H}){\rm exp}[-(400/T)^{2}]+1.4y({\rm H_2}){\rm
    exp}[-12000/(T+1200)]}~{\rm cm^{-3}}.
\end{equation}
Collisional dissociation and ionization absorb the same energy as the
binding energy, that is 4.48 eV per H$_2$ dissociation, 13.6 eV per H
ionization, 24.6 eV per He ionization, and 54.4 eV per He$^{+}$
ionization.

\subsection{Chemical reactions}
We solve nonequilibrium chemistry involving the four elements H, He, C and
O, that contains the following 45 species: 
${\rm H}$, ${\rm H_2}$, ${\rm e^-}$, ${\rm H^+}$, ${\rm H_2^+}$, 
${\rm H_3^+}$, ${\rm H^-}$, ${\rm He}$, ${\rm He^+}$, ${\rm He^{++}}$, 
${\rm HeH^{+}}$, ${\rm C}$, ${\rm C_2}$, ${\rm CH}$, ${\rm CH_2}$, 
${\rm CH_3}$, ${\rm CH_4}$, ${\rm C^+}$, ${\rm C_2^+}$, ${\rm CH^+}$, 
${\rm CH_2^+}$, ${\rm CH_3^+}$, ${\rm CH_4^+}$, ${\rm CH_5^+}$, ${\rm O}$, 
${\rm O_2}$, ${\rm OH}$, ${\rm CO}$, ${\rm H_2O}$, ${\rm HCO}$, 
${\rm O_2H}$, ${\rm CO_2}$, ${\rm H_2CO}$, ${\rm H_2O_2}$, ${\rm O^+}$, 
${\rm O_2^+}$, ${\rm OH^+}$, ${\rm CO^+}$, ${\rm H_2O^+}$, ${\rm HCO^+}$, 
${\rm O_2H^+}$, ${\rm H_3O^+}$, ${\rm H_2CO^+}$, ${\rm HCO_2^+}$ and
${\rm H_3CO^+}$.  
Chemical reactions included are listed in Appendix.
The reactions of H and He chemistry are mainly taken from Abel et
al. (1997) and Galli \& Palla (1998).
Other important H$_2$ forming processes included are the three-body
reactions (Palla et al. 1983) and the reaction on the surfaces of the
dust grains (Tielens \& Hollenbach 1985).  
In addition, we supplement H, He, C and O chemical reactions involving
the above species from Millar, Farquhar, \& Willacy (1997).    
\section{Results}
In this section, we present the results obtained by the method described 
in \S 2 and discuss our analysis.

Figure 1 displays the evolutionary trajectories of protostellar clouds whose 
metallicities are $z=0, 10^{-6}, 10^{-4}, 10^{-2}, 1$.
Here we set the initial condition to be $T=100 {\rm K}, n=1 {\rm
  cm^{-3}}, y(e)=1 \times 10^{-4}, y({\rm H_2})=1 \times 10^{-6}$.
All the carbon is assumed to be in the form of CII, while 
oxygen is OI at the beginning.  

The evolutionary trajectories for two other initial conditions are shown in 
Figure 2, as well as the same curves in Figure 1.
We can see from Figure 2 that trajectories of clouds with a fixed
composition converge rapidly toward the dense region for any initial
conditions (e.g., Hayashi \& Nakano 1965; Low \& Lynden-Bell 1976).
We have also tested the sensitivity to initial chemical compostions
for two cases -- (i) where the ionization degree $y(e)=1 \times 10^{-3}$
and (ii) all the carbon is assumed to be CI,-- and have found the
results similar to those shown in Figure 2.
Hereafter, we discuss only on the case presented in Figure 1.

In general, as seen in Figure 1, the temperatures of lower
metallicity clouds are higher because of their lower radiative cooling
ability (i.e., lower radiative cooling rate for the same temperature and
density) as long as the clouds are transparent to continuum and 
continue to collapse dynamically, owing to the efficient radiative
cooling, in other words, during what is known as the ``first collapse''
stage.  

On account of the total lack of metals and grains, the only cooling
agent in the temperature range below $\sim 10^{4}$K for primordial ($z=0$)
clouds is rovibrational transitions of molecular hydrogen (e.g., Matsuda 
et al. 1969).  
In the primordial gas, H$_2$ is formed mainly by the H$^{-}$ process,
\begin{eqnarray}
\label{eq:Hm1}
{\rm H}+e^{-} &\rightarrow & {\rm H^{-}}+\gamma ; \\
\label{eq:Hm2}
{\rm H}+{\rm H^{-}} &\rightarrow & {\rm H_{2}}+e^{-},
\end{eqnarray}
(Saslaw \& Zipoy 1967) until the density reaches $\sim
10^{8} {\rm cm^{-3}}$, where the three-body reactions  
\begin{equation}
\label{eq:3b1}
{\rm 3 H \rightarrow   H_2    +   H}
\end{equation}            
\begin{equation}
\label{eq:3b2}
{\rm 2 H     +   H_2    \rightarrow 2 H_2}
\end{equation}        
become efficient (Palla et al. 1983). 
H$_2$ line emission via electric quadrupole transitions
is the only efficient cooling agent until the number density exceeds
$\sim 10^{14} 
{\rm cm^{-3}}$, 
where H$_2$ continuum emission via collision-induced dipole transitions
begins to overwhelm. 
The drop in temperature at $n \simeq 10^{16} {\rm cm^{-3}}$ is
due to the quadrature dependence of the collision-induced emission 
coefficient on the density. 
The cloud becomes optically thick to the continuum at $10^{16} {\rm
  cm^{-3}}$. (See Fig. 3 a)

The gravitational contraction of metal-free protostellar clouds has been
investigated with Omukai \& Nishi (1998) by detailed hydrodynamical
calculations assuming spherical symmetry.  
Our evolutionary trajectory for the primordial cloud agrees well with
those of Omukai \& Nishi (1998) at the number density $n\gtrsim 10^{10}
{\rm cm^{-3}}$.
At lower densities, the temperature of ours is higher owing to the
influence of Omukai \& Nishi's (1998) initial condition.
They started the calculation from a cloud in hydrostatic equilibrium at $n
\simeq 10^{6} {\rm cm^{-3}}$. 
As long as the density is not so much higher than the initial
value, the collapse is slower than the free-fall rate, which we have 
assumed in this paper.
As a result, the temperature is lower because of the lower compressional
heating. 

The evolution of the cloud with $z=10^{-6}$ is essentially the
same as that of the primordial cloud, although there is minor deviation
caused by extra cooling owing to water molecules and dust grains.
(See Fig. 3 b) 
The dust thermal emission dominates the cooling just before the H$_2$
continuum becomes effective. 
However, it is only temporary and the grains rapidly sublimate at $n
\sim 10^{14} {\rm cm^{-3}}$. 
The most refractory grain composition that contributes to the Planck
mean opacity is iron, which
sublimate at about 1200K at this density (Table 3 of Pollack et
al. 1994).    
The slight amount of dust grains in the cloud with $z=10^{-6}$ does
not contribute to H$_2$ formation significantly, as seen in Figure 4.

For the cloud with $z=10^{-4}$, H$_2$, which is formed mainly on the
grain surfaces instead of the H$^{-}$ process, is the major coolant at
low densities ($<10^{4}{\rm cm^{-3}}$).
Water molecules, which are mainly formed at $\sim 10^{6} {\rm cm^{-3}}$
 by the reaction
\begin{equation}
{\rm H_2    +   OH    \rightarrow   H_2O   +   H}, 
\end{equation}
dominate the cooling in the range $10^{5} {\rm cm^{-3}}<
n < 10^{10} {\rm cm^{-3}}$.
In present-day molecular cloud cores, usually neither H$_2$O nor OH
plays a key role in the thermal balance since the temperature there is lower
than the excitation energy of these molecular species (Neufeld, Lepp, \&
Melnick 1995).   
However, in lower metal environments, the temperature is higher because
of the lower radiative cooling.
In that case, H$_2$O can be a principal cooling agent. 
In the cloud with $z=10^{-4}$, grain surface reactions exceed
the H$^{-}$ process as a mode of H$_2$ formation, although hydrogen does not
become fully molecular until the three-body reaction begins to work.
At $10^{10}{\rm cm^{-3}}$, the grain thermal emission dominates the cooling
and causes the temperature to drop to about 100K. 
The cloud becomes optically thick to the grain emission at about
$10^{13}{\rm cm^{-3}}$. (See Fig. 3 c)

For the cloud with $z>10^{-2}$, cooling is dominated by atomic
lines and CO lines in lower densities (that is, $n<10^{6} {\rm
  cm^{-3}}$ for $z=10^{-2}$, and $n<10^{3} {\rm cm^{-3}}$ for
$z=1$, respectively), and by the dust thermal emission in higher
densities. (See Fig. 3 d, e)  
The temperature of our $z=1$ cloud drops far less than 10K, which 
is appropriate in present-day protostellar clouds.
This is merely because we have ignored external radiation, and the
results should not be taken seriously as a model of present-day
molecular cloud cores. 
 
Clouds with metallicity $z>10^{-4}$ become
opaque to the thermal radiation of dust grains before the grain
vaporization occurs.  
Once the cloud becomes opaque, it begins to contract adiabatically since 
the radiative cooling rate drops rapidly.
According to some hydrodynamical simulations (e.g, Larson 1969; Masunaga et
al. 1998), a transient core in hydrostatic equilibrium (called a ``first
core'') forms after the cloud has more contractions.  
The hydrostatic core continues the adiabatic contraction by accreting the
envelope matter until the temperature reaches $\sim 2000$ K, where
molecular hydrogen begins to dissociate. 

Clouds with lower metallicity (i.e., $z<10^{-6}$) become opaque to the
H$_2$ continuum instead of the dust thermal emission, as described
above.
Since the temperature at that time is already near the dissociation
value, molecular hydrogen begins to dissociate just after the cloud
becomes opaque. 
As a result, no transient hydrostatic core forms in these cases as has
been demonstrated by the hydrodynamical calculations of Omukai \& Nishi
(1998).

When the ratio of specific heats $\Gamma={\rm dlog}p/{\rm dlog}\rho$
exceeds the critical value 4/3, corresponding to a gradient of
1/3 in the density-temperature ($n-T$) plane (i.e., Figure 1),
the bulk motion of the cloud is decelerated and
the free-fall assumption is no longer valid. 
However, the trajectories in the $n-T$ plane are not
altered since the clouds contract almost adiabatically anyway in the
opaque stage because of the rapidly declining radiative cooling rate
(e.g., Larson 1969; Narita, Nakano, \& Hayashi 1970).

In the course of the adiabatic contraction of the transient cores,
all the evolutionary trajectories converge to a certain line (Fig. 1,
{\it dashed line}) in spite of their different composition and
histories.  
We discuss the reason for the convergence in the following.
The trajectories in the $n-T$ plane just prior to when the clouds become
opaque are determined by thermal balance between the compressional
heating and the radiative cooling.
The compressional heating rate (per unit mass) is given by
\begin{equation}
\label{eq:Gcomp}
{\cal G}_{\rm comp}=-p \frac{d}{dt} (\frac{1}{\rho})=\frac{c_{\rm
    s}^2}{t_{\rm ff}},
\end{equation}
where $p$ is the pressure and $c_{\rm s}$ is the sound speed in the
central region of the cloud.
The radiative cooling at that time is dominated by continuum for all 
clouds, namely, the dust thermal emission for clouds with relative
metallicity $z>10^{-4}$, or the H$_2$ continuum emission for those with
lower metallicity.  
The radiative cooling rate is given by  
\begin{equation}
\label{eq:Lrad}
{\cal L}_{\rm rad}=4 \kappa_{\rm B} \sigma T^{4},
\end{equation}
where $\kappa_{\rm B}$ is the Planck mean continuum opacity.
In the equation above, we have used the relation that the grain temperature is 
the same as the gas temperature, $T_{\rm gr}=T$, which is valid in a
cloud so dense that it becomes opaque to the continuum.
Note that expression (\ref{eq:Lrad}) holds for any clouds that cool 
mainly by continuum by choosing an appropriate value of opacity, where
the opacity
is an increasing function of relative metallicity $z$.
Equating (\ref{eq:Gcomp}) and (\ref{eq:Lrad}), for each value of 
$\kappa_{\rm B}(z)$, we obtain a trajectory just before the cloud becomes
opaque
(we call it Tr($\kappa_{\rm B}$)); 
\begin{equation}
\label{eq:GLpath}
{\rm Tr(\kappa_{\rm B})}:~~~\frac{c_{\rm s}^2}{t_{\rm ff}}=4 \kappa_{\rm B}
\sigma T^{4}.  
\end{equation}
As the cloud of opacity $\kappa_{\rm B}$ collapses along the trajectory
represented by (\ref{eq:GLpath}), the central part of the cloud
becomes opaque to the continuum at the point 
\begin{equation}
\label{eq:P}
{\rm P(\kappa_{\rm B})}:~~ \tau_{\rm J}(\kappa_{\rm B},n,T) =\kappa_{\rm
    B} \rho \pi \frac{c_{\rm s}}{\sqrt{G \rho}}=1,
\end{equation}
where we have used the assumption that the size of the central region is 
approximately $\lambda_{\rm J}$.
Eliminating $\kappa_{\rm B}$ in equations (\ref{eq:GLpath}) and
(\ref{eq:P}), we obtain the locus L of points P for variable $\kappa_{\rm B}$;
\begin{equation}
\label{eq:tauj}
{\rm L}:~~T=(\frac{k^3}{12 \sigma ^2 m_{\rm p}})^{1/5} n^{2/5}. 
\end{equation}
On the other hand, the trajectory of the cloud of opacity
$\kappa_{\rm B}$ in the opaque stage, O($\kappa_{\rm B}$), is the line
that starts from the point P($\kappa_{\rm B}$) and whose gradient is
approximately equal to 
the value for adiabatic contraction $\gamma_{\rm ad}-1$.
The adiabatic exponent $\gamma_{\rm ad}$ for molecular hydrogen is 7/5
in the temperature range $T\gtrsim 200$K, while it is 2/3 in the lowest
temperature.
The gradient of the trajectory O($\kappa_{\rm B}$) in the range
$T\gtrsim 200$K is 2/5.
This value happens to be the same as the gradient of the line L.     
Thus, the trajectories in the opaque stage O all coincide with the
line L, which can be written only by physical constants. 
When the clouds become opaque (i.e., at the point P),
their specific entropy has the same value for all clouds regardless of
their metallicity.  

As a cloud climbs up along line L, the temperature  
reaches the dissociation value ($\simeq 2000$K) at $n\simeq 3\times
10^{16}{\rm cm^{-3}}$.   
The ratio of specific heats are reduced below the critical value 4/3.
Then the cloud begins to collapse dynamically again, or in other words, it
begins the second collapse (e.g., Larson 1969). 
The trajectories in the second collapse stage are also unique because the
specific entropy has the same value for all clouds regardless of their
metallicity.  
When most hydrogen molecules have been dissociated, the ratio of specific
heats rises above the critical value 4/3 again.
After the central part of the cloud contracts almost adiabatically to
some extent, a hydrostatic core forms.
This core is referred to as the stellar or second core in literature.
We cannot accurately estimate the size of the stellar core from our
simple one-zone treatment.
However, hydrodynamical calculations tell us that the number density and
mass of the stellar core at its formation time are on the order of $10^{22}
{\rm cm^{-3}}$ and $10^{-3} M_{\sun}$ respectively, both for present-day
(Larson 1969) and primordial protostars (Omukai \& Nishi 1998). 
This value is expected to be universal, namely, independent of
metallicity since the equations of state, or trajectories in the $n-T$ 
plane, are the same in the second collapse stage. 

The stellar core, or protostar, grows in mass by accretion of the
envelope material. 
The temperature rises accordingly and eventually the core will
become an ordinary star. 
Consequently, the final mass of stars is determined by the subsequent
accretion onto the core, although the size of the protostar at the time of
formation is the same. 
Stahler, Shu, \& Taam (1980) argued that a rough estimate of the
protostellar mass accretion rate ${\dot M}$ can be obtained from the
relation 
\begin{equation}
{\dot M} \sim c_{\rm s}^{3}/G, 
\end{equation}
where $c_{\rm s}$ is the isothermal sound speed in the initial
protostellar cloud.
Therefore, the mass accretion rate is higher for a lower
metallicity cloud because of the higher temperature of the protostellar
cloud (Stahler, Palla, \& Salpeter 1986).
The higher accretion rate and lower opacity (then smaller radiation
force) may result in the higher mass of formed stars for a lower
metallicity cloud (e.g., Wolfire \& Cassinelli 1987). 
However, to address these issues fully, further investigations are
needed.
\section{Summary}
We have investigated the thermal and chemical evolution of collapsing
protostellar clouds with different metallicities.
The varied range of metallicity spans the local interstellar value
($\simeq 1 Z_{\sun}$) to zero.

The evolution of the clouds is summarized as follows.
While the clouds are transparent to continuous radiation,  
the temperature of clouds with lower metallicity is higher 
since their radiative cooling rates are lower for the same density and
temperature.
However, after the clouds become opaque and begin adiabatic
contraction, their evolutionary trajectories
converge to a line that is determined only by physical constants.
Thereafter, the trajectories coincide with each other regardless of their
metallicity.
Consequently, the physical dimension of the stellar core at the time of
formation is the same for clouds with any composition.

We have also discussed analytically the reasons for the convergence.

\acknowledgements
The author wishes to express his cordial thanks to Humitaka Sato and
Naoshi Sugiyama for their continual interest and advice, to Hiroshi
Koyama, Ryoichi Nishi, and Toru Tsuribe for useful discussions, and to
the referee for improving this manuscript.
\newpage



\newpage

\begin{figure} 
\plotone{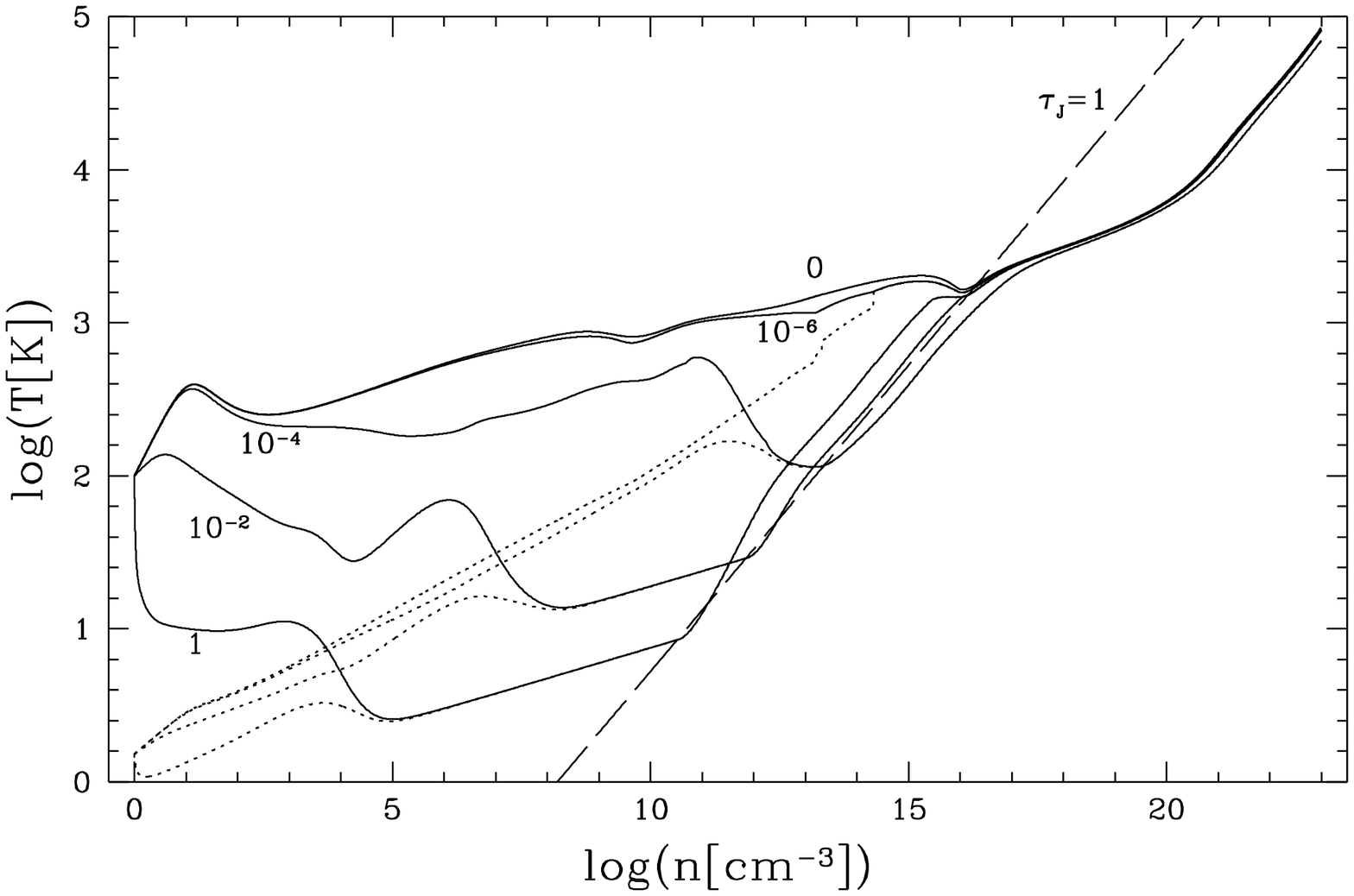}
\caption{The evolutionary trajectories of protostellar clouds with
  different metallicities. 
The solid curves show the temperature of the clouds with metallicity
({\it from the top to bottom}) $z=0, 10^{-6}, 10^{-4}, 10^{-2}, 1$.
The dotted curves show the grain temperature for the same clouds. 
The initial condition is $T=100 {\rm K}, n=1 {\rm
  cm^{-3}}, y(e)=1 \times 10^{-4}, y({\rm H_2})=1 \times 10^{-6} $.
Initially, all the carbon is assumed to be in the form of CII, while 
oxygen is OI.
The dashed line marked ``$\tau_{\rm J}=1$'' represents the line L in the
text (Eq.35). 
\label{fig1}}
\end{figure}

\begin{figure} 
\plotone{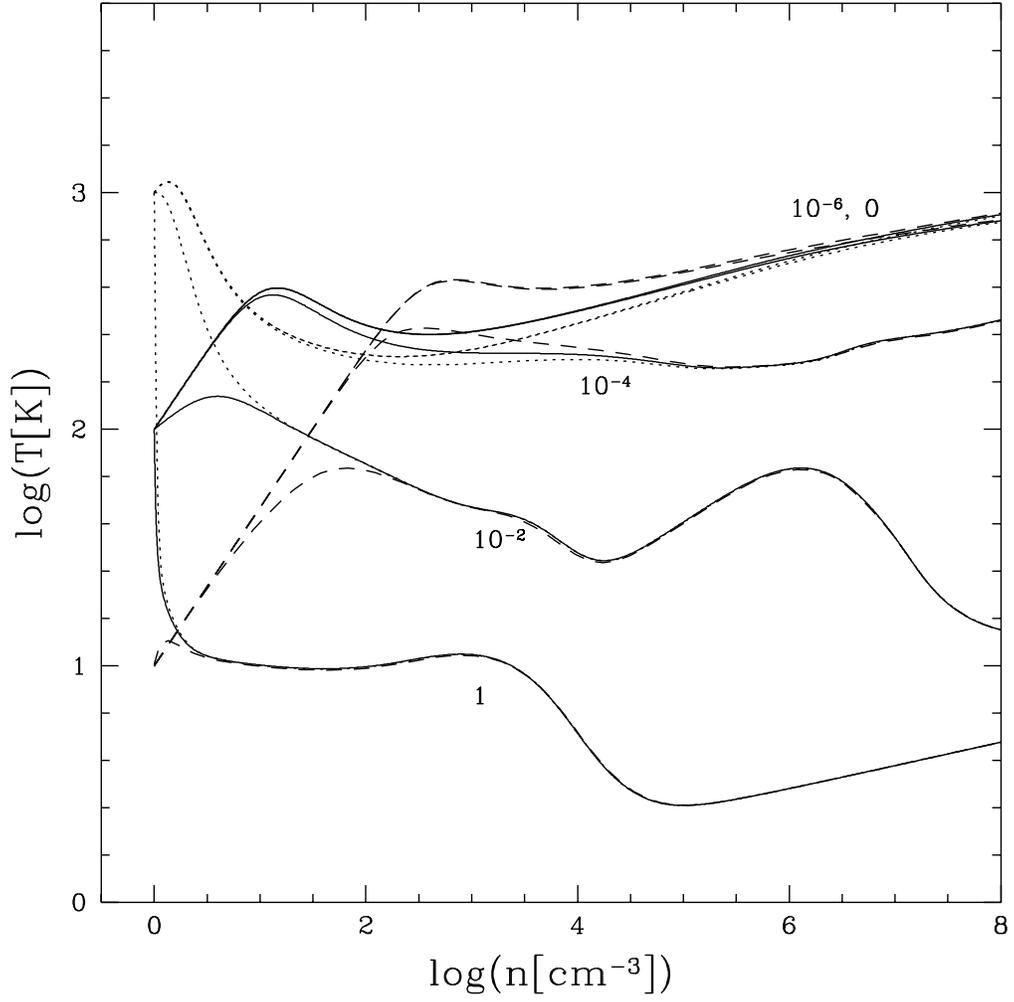}
\caption{Initial condition dependence of the evolutionary
  trajectories. The solid curves are the same as those of Figure 1
  (``fiducial model''). 
The initial temperature of dotted curves is 1000K and of the dashed
  curves is 10K, while other quatities are the same as the fiducial model.
For same values of metallicity, a wide variation in temperature yields
convergent evolutionary tracks.
\label{fig2}}
\end{figure}
\begin{figure} 
\plotone{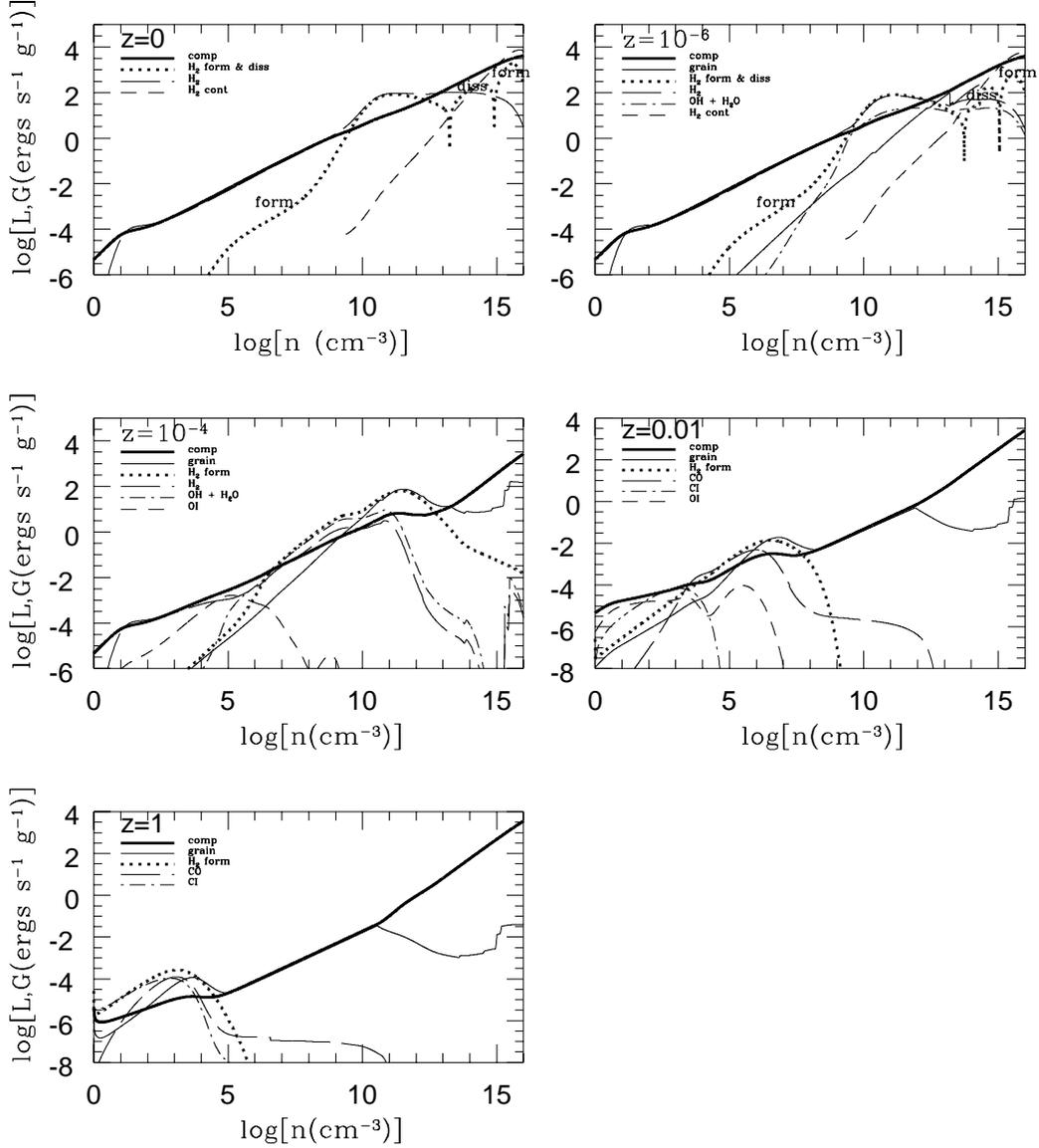}
\caption{The contributions to the cooling/heating rates by major processes
  for the same clouds as illustrated in Figure 1: (a)$z=0$, (b)$z=10^{-6}$,
  (c)$z=10^{-4}$, (d)$z=10^{-2}$, and (e)$z=1$.
  Curves labelled ``comp'', ``grain'', ``H$_2$ form/diss'', and ``H$_2$
  cont'' show the cooling/heating rate by the adiabatic compression, the
  gas-grain heat transfer, formation/dissociation of H$_2$, and H$_2$
  continuum. 
  The other curves labelled  ``H$_2$'', ``OH+H$_2$O'', ``CO'', ``CI'',
  and ``OI'' illustrate the line cooling rate by these atoms/molecules.
  Irregularities in ``grain'' (in panels b-e) and
  ``H$_2$'' and ``OH+H$_2$O''(in panel b and c) cooling rates near $n \sim 
  10^{14} {\rm cm^{-3}}$ are caused by sublimaiton of each grain
  component and subsequent change of the dust opacity.    
\label{fig3}}
\end{figure}

\begin{figure} 
\plotone{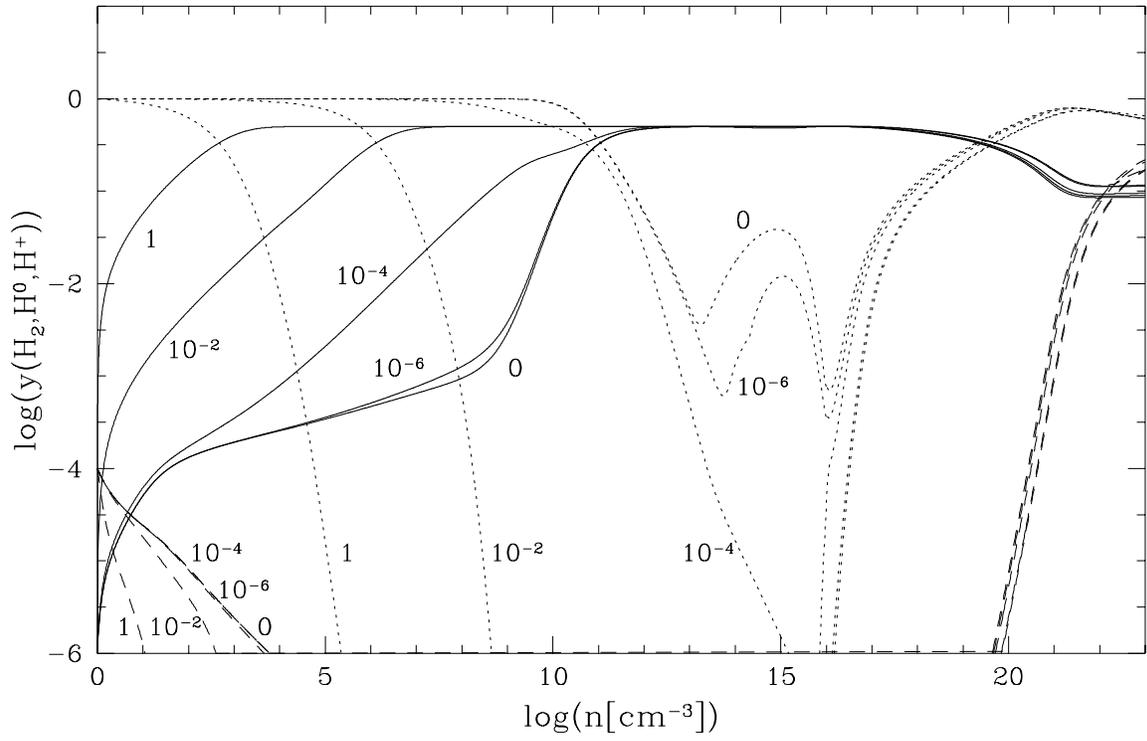}
\caption{The concentrations of hydrogen molecules ({\it solid line}), atoms
  ({\it dotted line}) and ions ({\it dashed line}) for the
  same clouds as illustrated in Fig.1.
\label{fig4}}
\end{figure}
\end{document}